\documentclass[a4paper,11pt]{article}
\usepackage{amssymb,amsmath,bm}
\usepackage{graphics,graphicx}
\usepackage{array,booktabs}
\usepackage{authblk}
\usepackage{cite}
\usepackage{xcolor}
\usepackage[margin=2.5cm]{geometry}
\usepackage{caption}
\usepackage{epstopdf}
\usepackage{subfig}
\usepackage{graphicx}
\usepackage{tikz-feynman}
\usepackage{tikz}
\usepackage{multirow}
\usepackage{pgfplots}
\usepackage[linktocpage,
colorlinks = true,
linkcolor = blue,
urlcolor  = blue,
citecolor = red,
anchorcolor = blue]{hyperref}
\usepackage{orcidlink}
\usepackage{epsfig}
\numberwithin{equation}{section}
\title{\textbf{Vector dark matter and LHC constraints, including a 95 GeV light Higgs boson}}
\author[a]{Seyed Yaser Ayazi\thanks{syaser.ayazi@semnan.ac.ir}\orcidlink{0000-0002-5994-3316}}
\author[a]{Mojtaba Hosseini\thanks{mojtaba\textunderscore hosseini@semnan.ac.ir}}
\author[b]{Saeid Paktinat Mehdiabadi\thanks{spaktinat@yazd.ac.ir}\orcidlink{0000-0001-7303-0217}}
\author[a]{Rouzbeh Rouzbehi\thanks{rouzbehrouzbehi@semnan.ac.ir}\orcidlink{0009-0005-7139-5841}}
\affil[a]{Physics Department, Semnan University, P.O. Box 35131-19111, Semnan, Iran}
\affil[b]{Department of Physics, Yazd University, P.O. Box 89195-741, Yazd, Iran}
\date{\today}

\begin{document}

\baselineskip 0.6 cm

\maketitle

\begin{abstract}
We study LHC searches for an extension of the Standard Model (SM) by exploiting an additional Abelian $U_D(1)$ gauge symmetry and a complex scalar Higgs portal. As the scalar is charged under this gauge symmetry, a vector dark matter (VDM) candidate can satisfy the observed relic
abundance and limits from direct dark matter (DM) searches. The ATLAS and CMS experiments have developed a broad search program for the DM candidates, including associate production of Higgs boson, $Z$ boson, and top quark that couple to DM. In this paper,  we perform an extensive analysis to constrain the model by using these experiments at LHC. It can be seen that the LHC results can exclude some parts of the parameter space that are still allowed by relic density and the direct detection searches.  Using the LHC results, all scalar Higgs portal masses are excluded for the light VDM. Furthermore, exclusion limits on the parameter space of the model by using the new results of the CMS and ATLAS Collaborations for a new light Higgs boson with mass $\sim95~\rm GeV$ are provided.

\end{abstract}

\section{Introduction} \label{sec:intro}
The existence of dark matter (DM) is largely known through a variety of astrophysical and cosmological experiments\cite{Bertone:2016nfn}. Although it is measured to be about $26.4\%$ of the total energy density of the Universe via indirect observations, the true nature of DM particles is still unknown. However, we know that DM must be electrically neutral and does not interact with ordinary baryonic matter.

The main effects caused by DM are gravitational, but the existence of a
weakly interacting massive particle (WIMP) is often hypothesised, since it leads to the correct relic density for nonrelativistic matter in the early Universe (WIMP miracle)\cite{Feng:2022rxt}. In general,
the dark sector can interact with the Standard Model (SM) particles through the Higgs portal \cite{Lebedev:2021xey}. We consider a $U(1)$ extension of SM in which a scalar mediated between SM and the dark sector and a vector can play the role of a DM particle. Experiments around the world seek to unravel the nature of DM with different strategies. One of these strategies is to search at particle colliders. Such dark particles could be produced in particle colliders and detected
as the large missing transverse momentum produced in association with the
SM particles. To have a realistic model, it must satisfy different available constraints from the DM abundance via standard thermal freeze-out processes, direct detection experiments, and the LHC searches for the DM particles. In this regard, we perform an analysis on the model, focusing on relic density, direct detection, and several searches on associate production of Higgs boson, $Z$ gauge boson, and top quark with the DM particle. In particular, we determine the region of parameter space that can be excluded by the searches at LHC. Some of the LHC analyses have reported the data and background event yields in different signal regions. It enables us to find the limits on the DM signal yields by some statistical methods. By implementing the signal selections in an event generator, one can find the  DM signal yields in different points of the parameter space. This yield is compared to the limits extracted from the LHC analyses to decide whether a point is allowed or excluded.

The recent results, reported by the CMS and ATLAS Collaborations, indicate an excess in the $\gamma$$\gamma$ and $\tau$$\tau$ final states consistent with a light Higgs boson ($\sim95~\rm GeV$) \cite{CMS:2024yhz}-\cite{ATLAS:2024bjr}. They provide interesting hints for new physics. Motivated by these results, we perform a study to constrain the model by considering a $95~\rm GeV$ light Higgs boson.

To explain the model and analysis, the paper is organized as follows. In Sec. \ref{sec.model}, we introduce the model under
consideration. In Secs. \ref{sec.relic} and \ref{sec.direct}, the allowed parameter space of the model is probed by considering the relic density of DM and direct detection experiments.  We present the method and the main results for the LHC constraints on the model by a list of processes in Sec. \ref{sec.lhc}. In Sec. \ref{sec.lhc95}, the excluded region in the light of the newly observed $95~\rm GeV$ light Higgs boson is reported.  We also study indirect constraint from the H.E.S.S experiment on the DM candidate in Sec. \ref{sec.hess}. Finally, Sec. \ref{sec.con} concludes the paper.

\section{The Model} \label{sec.model}
For the reader’s convenience, we briefly introduce the main characteristics of the model that is fully developed in Ref.\cite{YaserAyazi:2022tbn}. We extend the SM  by a dark sector where the $U_D(1)$ Abelian gauge symmetry is spontaneously broken by the dark Higgs mechanism. We introduce a complex
scalar $S$ which has a unit charge under $U_D(1)$ and the $U_D(1)$ gauge field  $V_{\mu}$.  These new fields are singlet under the SM gauge group. We also consider an additional $Z_2$ symmetry, under which the vector field $V_{\mu}$ and the scalar field $S$ transform as follows:
\begin{equation}
V_{\mu}\rightarrow-V_{\mu},~~~ S\rightarrow S^*
\end{equation}

Because of the imposed $Z_2$ symmetry under which only the dark gauge
boson $V_{\mu}$  is odd, kinetic mixing between the gauge bosons of the $U_D(1)$ and to $U(1)_Y$ is not
possible and two sectors will not mix at any order of perturbation theory; therefore, the field
renormalization constants are defined independently in each sector\cite{Glaus:2019itb,Hashino:2018zsi,Amiri:2022cbv}. Therefore, the vector field $V_{\mu}$ can be considered as a DM candidate. The total Lagrangian of the model is given by

\begin{equation}
 {\cal L} ={\cal L}_{SM}- \frac{1}{4} V_{\mu \nu} V^{\mu \nu}+ (D'_{\mu} S)^{*} (D'^{\mu} S) - V(H,S)  , \label{2-2}
\end{equation}
where $ {\cal L} _{SM} $ is the SM Lagrangian without the Higgs potential term and the covariant derivative and field strength of $V_{\mu}$ are given as
\begin{align}
& D'_{\mu} S= (\partial_{\mu} + i g_v V_{\mu}) S,\nonumber \\
& V_{\mu \nu}= \partial_{\mu} V_{\nu} - \partial_{\nu} V_{\mu}.\nonumber \end{align}
The potential, which is renormalizable and invariant
under gauge and $ Z_{2} $ symmetry, is
\begin{equation}
V(H,S) = -\mu_{H}^2 H^{\dagger}H-\mu_{S}^2 S^*S+\lambda_{H} (H^{\dagger}H)^{2} + \lambda_{S} (S^*S)^{2} +  \lambda_{S H} (S^*S) (H^{\dagger}H). \label{2-3}
\end{equation}
The quartic interaction, $ \lambda_{SH} (S^*S) (H^{\dagger}H) $, is the only connection between the dark sector and the SM. The SM Higgs field as well as dark scalar field can receive vacuum expectation values, breaking  the electroweak and $U_D(1)$ symmetries, respectively. In this gauge, we write down the
Higgs field as
\begin{equation}
	H = \frac{1}{\sqrt{2}} \left( \begin{array}{c}
		0  \\
		\nu_{1}+h_1
	\end{array} \right)\,
\end{equation}
and $S$ as \begin{equation}
	S=  \nu_{2} + h_2\,.
\end{equation}
$\nu _1$ and $\nu _2$ are vacuum expectation values of Higgs fields in which we suppose $\nu_{1}$ = 246 GeV. After the electroweak symmetry breaking \cite{YaserAyazi:2022tbn}, the parameters of the model are given by
 \begin{align}
& \nu _2=\frac{M_V}{g_v} \nonumber,~~~~~~~~~~ \sin\alpha=\frac{\nu_1}{\sqrt{\nu _1^2+\nu_2^2}}, \\
& \lambda _H=\frac{\cos ^2\alpha M_{H_1}^2+\sin ^2\alpha  M_{H_2}^2}{2 \nu _1^2},  \nonumber \\
& \lambda _S=\frac{\sin ^2\alpha M_{H_1}^2+\cos ^2\alpha  M_{H_2}^2}{2 \nu _2^2},  \nonumber \\
&  \lambda _{SH}=\frac{ \left(M_{H_2}^2-M_{H_1}^2\right) \sin \alpha  \cos \alpha}{\nu _1 \nu _2}, \label{cons}
\end{align}
where $\alpha$ is the mixing angle between weak eigenstates and mass eigenstates. 
The mass eigenstates of Higgs fields can be written as follows:
\begin{align}
M^2_{H_{2},H_{1}}=\lambda_H \nu_1^2+\lambda_S \nu_2^2 \pm \sqrt{(\lambda_H \nu_1^2-\lambda_S \nu_2^2)^2+\lambda_{SH}^2\nu_1^2\nu_2^2},
\label{mas}
\end{align}
where we suppose $ M_{H_1} = 125~\rm  GeV$. It is remarkable that this construction is rather minimal, in the sense that only three new free parameters are
added: coupling $ g_v $ and new mass parameters $ M_{H_2} $ and $M_V $.

 We also have studied various theoretical and experimental constraints on the parameter space of the model. For theoretical constraints, we must consider constraints such as  perturbativity, unitarity, and positivity of the potential. In this regard, we suppose all theoretical  conditions that have been discussed in \cite{YaserAyazi:2022tbn}. For experimental constraints, we consider invisible Higgs decay, relic density, and direct detection bounds. The constraints for invisible Higgs decay are completely similar to \cite{YaserAyazi:2022tbn} but for two other constraints, we extend the studied parameter space.

\section{Relic density} \label{sec.relic}
\begin{figure}
	\begin{center}
		\centerline{\hspace{0cm}\epsfig{figure=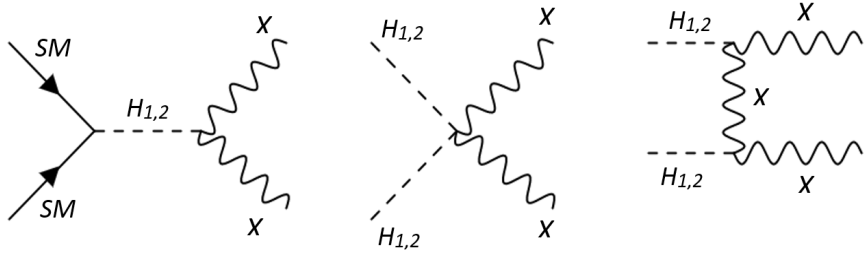,width=12cm}}
		\centerline{\vspace{-0.2cm}}
		\caption{The dominant Feynman diagrams for DM relic density production cross section.} \label{feynman}
	\end{center}
\end{figure}

\begin{figure}
	\begin{center}
		\centerline{\hspace{0cm}\epsfig{figure=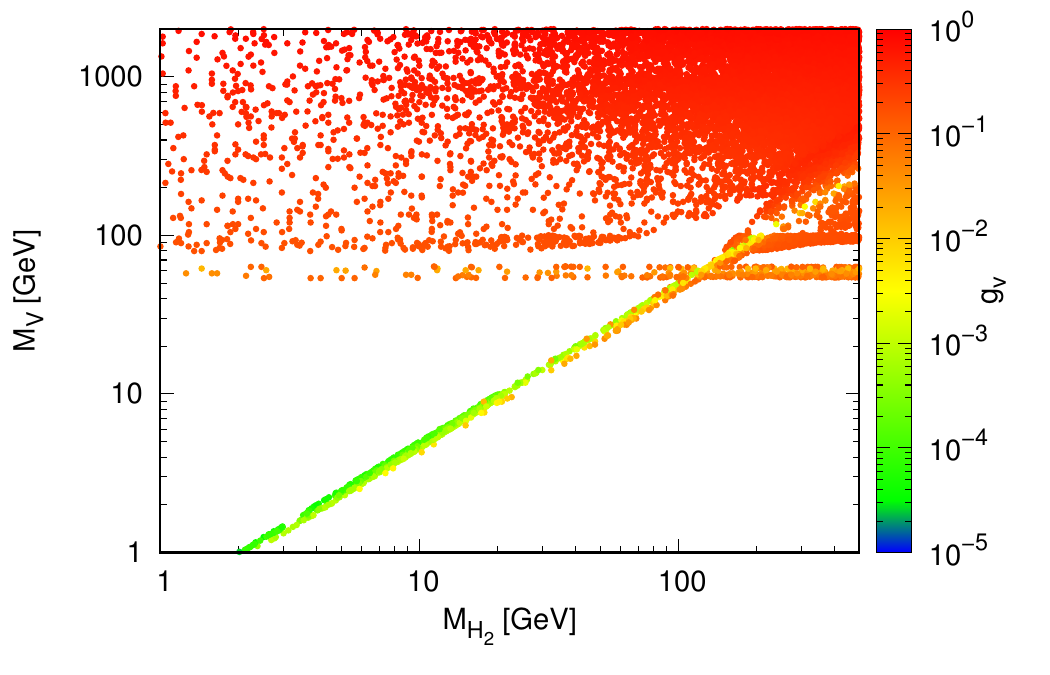,width=12cm}}

		\caption{The allowed range of parameter space consistent with the DM relic density.} \label{Relic}
	\end{center}
\end{figure}

 The evolution of the number density of DM particles $(n_X)$ with time is governed by the Boltzmann equation:
\begin{equation} \label{44}
\dot{n}_X + 3Hn_X = -\langle\sigma_{ann} \nu_{rel}\rangle [n_X ^2 -(n_X ^{eq})^2] ,
\end{equation}
where $H$ is the Hubble parameter and $n_X ^{eq} \sim (m_X T)^{3/2} e^{-m_X /T} $ is the particle density before particles get out of equilibrium. The dominant Feynman diagrams for DM production processes are shown in Fig.~\ref{feynman}.
In this regard, we calculate the relic density
numerically for the vector dark matter (VDM) particle by implementing  the model into micrOMEGAs \cite{Barducci:2016pcb}. The allowed range of parameter space corresponding to observed DM relic density( according to the data of the Planck Collaboration \cite{Planck:2018vyg}) is depicted in Fig.~\ref{Relic}. As can be seen, there are three mass regions with the correct relic density: (i) $M_V\sim M_{H_2}/2$ , (ii) $M_V\lesssim M_{H_2}$, and (iii) $M_V > M_{H_2}$\cite{Amiri:2022cbv}. In the first
case, what is clear is that DM annihilates resonantly via $H_2$, then the latter decays into SM states(the strip in the figure), and a small mixing angle($\alpha$) makes the resonant s-channel annihilation less effective.  In the second case, $VV \rightarrow H_2 H_2$ annihilation becomes relevant. Of course when $M_V < M_{H_2}$, due to the thermal tail with high velocity, the correct relic density can be obtained from Ref.~\cite{DAgnolo:2015ujb}. In the third case, DM annihilates into lighter $H_2$ states. But what is very clear is that the mass window for the resonance and forbidden region(the first and second cases) is quite limited, whereas the case of $M_V > M_{H_2}$ involves much more parameter space. So, the most allowed points of the model are in the region of $60  <  M_V < 2000 \rm ~GeV$ and $ 0.1 < g_v < 1$. Of course, for $ M_V <60 \rm ~GeV$ the sin $\alpha<0.44$ constraint is also one of the reasons that limited the parameter space\cite{Farzinnia:2013pga,Farzinnia:2014xia}.

\section{Direct Detection} \label{sec.direct}
\begin{figure}
	\begin{center}
		\centerline{\hspace{0cm}\epsfig{figure=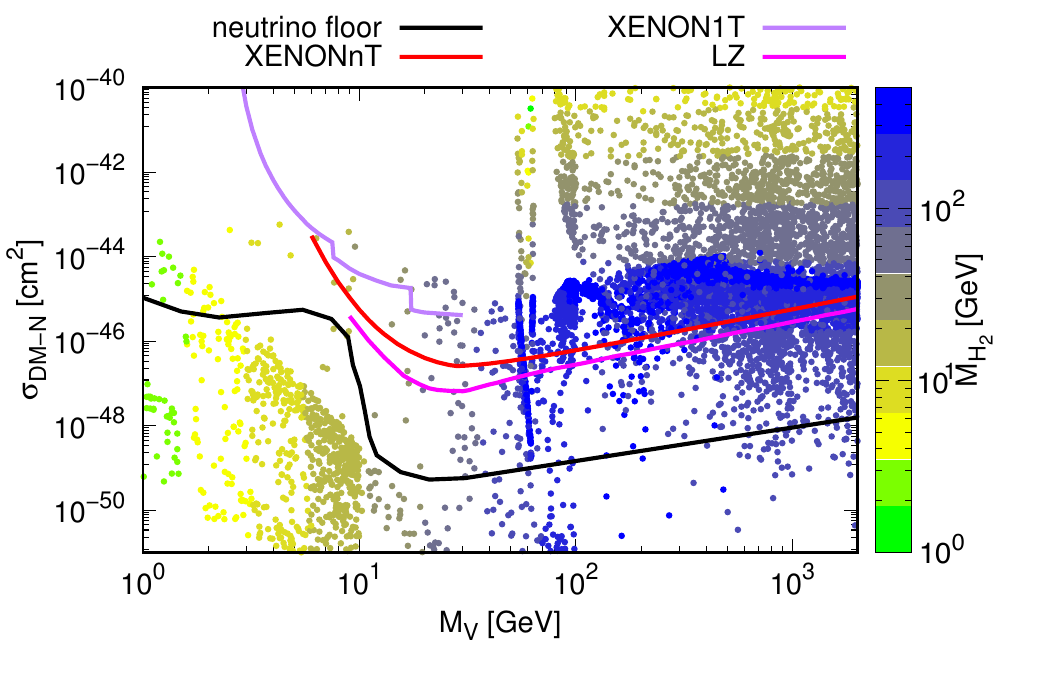,width=12cm}}
		\centerline{\vspace{-0.2cm}}
		\caption{The allowed range of parameter space consistent with the DM relic density and DD.} \label{direct detection}
	\end{center}
\end{figure}

 Before examining the constraints of LHC on the various channels, let us turn our attention to the direct detection(DD) of VDM in the model. At the tree level, a VDM particle can interact elastically with a nucleon either through $H_1$ or via $H_2$ exchange\cite{YaserAyazi:2019caf}. We use the XENONnT and LUX-ZEPLIN(LZ) experiments results to constrain the parameter space\cite{XENON:2023cxc,LZ:2022lsv}. Also for $M_V<30~\rm GeV$, the results of the XENON1T experiment are used\cite{XENON:2019gfn}. Figure \ref{direct detection} shows the parameter space of the model in agreement with the DD and relic density constraints. As it can be seen, for different values of the DM mass, there are points that are protected from all the presented phenomenological constraints.

 It is necessary to mention that since we are motivated to follow DM phenomenology at the LHC, we consider our parameter space at the electroweak range (1--2000 GeV). In this regard, we used the micrOMEGAs package to calculate the relic density and direct detection constraints. We set $1\times 10^6$ points randomly and obtained those consistent with the allowed relic density and direct detection bound. The final results are shown in Figs. \ref{Relic} and \ref{direct detection}.

\section{The LHC searches and the DM constraints} \label{sec.lhc}
The LHC experiments at CERN have done a long list of searches for new physics beyond the SM (BSM) and also measurements of SM cross sections and variables. Up to this date, no sign of new physics or significant deviation from the SM predictions has been reported. This lack of experimental evidence for new physics can be used to constrain any newly introduced extension of SM, e.g., the proposed DM model. To produce the results of this section, the event yields and uncertainties reported by the LHC experiments are used to find the 95\% and 68\% confidence level (CL) upper limit on the event yields of BSM. The procedure is done using the tools provided in ROOT~\cite{Brun:1997pa} analysis framework. It is based on a likelihood ratio semi-bayesian method~\cite{Cousins:1991qz}-\cite{Conrad:2002kn}. The idea is to find the maximum room for events from BSM, by comparing the expected number of events from the SM and the observed event yields in data.  The authors have used a similar method to constrain other physics BSMs in  Ref.~\cite{PaktinatMehdiabadi:2017clu}.

To find the upper limits, 20\% total statistical and systematic uncertainty is assumed on DM signals. To make sure the results do not depend on this value, the values of 15\% and 25\% are also used for this total uncertainty and the changes in the results are found to be marginal. The found maximum number of BSM events is divided by the corresponding integrated luminosity to find the upper limit on the cross section times the acceptance times the experimental efficiency of the BSM in question. To continue, this quantity is called the upper limit of the visible  cross section ($\sigma^{UL}_{vis}$). For various channels $\sigma^{UL}_{vis}$ values are shown in Table~\ref{r80}.\\
\begin{table}
\centering
\caption{The upper limit of the visible $pp$ cross section ($\sigma^{UL}_{vis}$) for different channels based on the LHC searches at $\sqrt s=13$ TeV}
\begin{tabular}{|c|c|c|c|c|}
\hline
\multirow{3}{*}{Channel} & \multicolumn{4}{c|}{ $\sigma^{UL}_{vis}$ (fb)}  \\\cline{2-5}
          & \multicolumn{2}{c|}{ CMS} & \multicolumn{2}{c|}{ ATLAS}\\\cline{2-5}
                   & 95\%CL & 68\%CL& 95\%CL & 68\%CL\\
\hline
$VVZ, Z \to \ell^+\ell^-$&1.22 & 0.66 &4.43 & 3.41\\ \hline
$H_1VV, H_1\to \tau^+\tau^-$      &0.75 & 0.39  & 0.06 & 0.03\\ \hline
$H_1VV, H_1\to b\bar{b}$      &1.75 & 0.92  & 2.6 & 1.36\\ \hline
$ t\bar{t}VV$&85.91&44.1& 0.66 &0.39\\\hline
\end{tabular}
\label{r80}
\end{table}
To generate the events MadGraph5aMC@NLOv.3.4.2~\cite{Alwall_2014} (MG) is used, but to make the model understandable by MG, the introduced model is implemented in LanHEP\cite{Semenov:2014rea}. The output of this package can be used in MG, easily. The analysis selections are implemented in MG to simulate the experimental acceptance as accurately as possible. The generated events are passed to PYTHIA8\cite{Bierlich:2022pfr} for patron showering and hadronization. The experimental efficiency within the acceptance is assumed to be 100\%, which is an optimistic assumption and ends up with conservative results in the shown plots. To have a more realistic estimation of the efficiencies, one needs to pass the generated events through the full detector simulation and reconstruct the events in detail, which is beyond the scope of this analysis. In most of the LHC analyses, there are several signal regions (bins). The event yields from SM prediction in different bins are summed up and treated as a single bin analysis. The same is done for the collision data, also. Special attention is paid to have completely separated bins and one event has no chance to appear in two different bins and be taken into account two times.  For each test point, the input parameters are set in MG and 10000 events are generated to measure the production cross section of that test point. The measured cross section by MG is compared with $\sigma^{UL}_{vis}$. Only a test point with a measured production cross section less than $\sigma^{UL}_{vis}$ is considered as an allowed point; otherwise, it is marked as excluded.
\subsection{LHC constraints on $H$+DM}
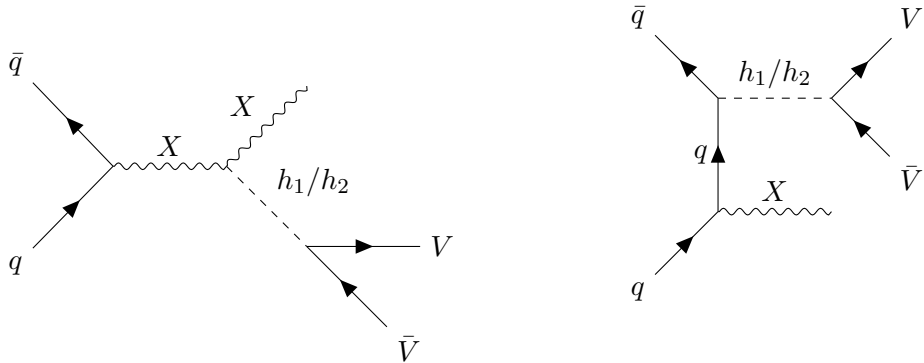
\begin{figure}[h]
	\centering
	\begin{tikzpicture}
		\begin{feynman}
			\vertex (a);
			\vertex [below left=of a] (b) {$q$};
			\vertex [above left=of a] (c) {$\bar{q}$};
			\vertex [right=of a] (d);
			\vertex [above right=of d] (e);
			\vertex [below right=of d] (h);
			\vertex [right=of h] (j) {$V$};
			\vertex [below right=of h] (i) {$\bar{V}$};
			\diagram{
				(b) -- [fermion] (a) -- [fermion] (c);
				(a) -- [boson, edge label=$X$] (d);
				(d) --[boson, edge label=$X$] (e);
				(d) -- [scalar, edge label=$h_1/h_2$] (h);
				(i) -- [fermion] (h) -- [fermion] (j);
			};
			\begin{scope}[xshift=8cm, yshift=-0.6cm]
				\vertex (a);
				\vertex [above=of a] (b);
				\vertex [below left of=a] (c) {$q$};
				\vertex [above left of=b] (d) {$\bar{q}$};
				\vertex [right of=a] (e);
				\vertex [right of=b] (f);
				\vertex [above right of=f] (g) {$V$};
				\vertex [below right of=f] (h) {$\bar{V}$};
				\diagram{
					(a) -- [fermion, edge label=$q$] (b);
					(c) -- [fermion] (a);
					(b) -- [fermion] (d);
					(a) --[boson, edge label=$X$] (e);
					(b) -- [scalar, edge label=$h_1/h_2$] (f);
					(h) -- [fermion] (f) -- [fermion] (g);
				};
			\end{scope}
		\end{feynman}
	\end{tikzpicture}
	\caption{Feynman diagrams relevant for the production of $X=Z/H_1$ in association with DM at the LHC in the process $pp \to XVV$.}
	\label{diag.XVV}
\end{figure}

Both multipurpose LHC experiments have reported results on the search for associate production of the Higgs boson and DM. Some of the leading order Feynman diagrams, contributing in this process, are shown in Fig.~\ref{diag.XVV}.
One of the most interesting decay modes of the Higgs boson is its decay to two photons via a top quark mediated loop. Since the proposed model is implemented in leading order, it cannot handle such decays and this final state is ignored in this analysis. To continue, the decays of the Higgs boson to two $\tau$ leptons or two b quarks are considered. The CMS experiment has reported the event yields of the SM prediction and collision data after applying the cuts in events with an SM Higgs boson and the missing transverse momentum in Ref.\cite{CMS:2018zjv}. The analysis uses $35.9~\rm fb^{-1}$ of data in $\sqrt{s}=13~\rm TeV$. The paper contains the results of Higgs boson decay to two b quarks. A special algorithm is used to identify the Higgs bosons with  medium Lorentz boost, where  b jets produced in the decay of the Higgs boson are merged and appear as a large-radius jet. The yields are reported in four nonoverlapping bins of the missing transverse momentum.
The ATLAS experiment has done a similar search using $139~\rm fb^{-1}$ of data in $\sqrt{s}=13~\rm TeV$ \cite{ATLAS:2021shl}. They also use a  variable-radius track-jet algorithm to treat the boosted Higgs bosons properly. For events with exactly two b jets, five exclusive bins of the missing transverse momentum are defined to report the yields.

\begin{figure}
	\centering
\subfloat[CMS]{\includegraphics[width=0.48\linewidth]{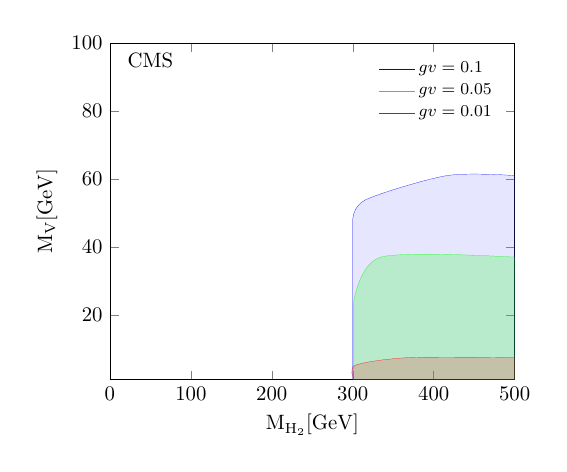}}
 \subfloat[ATLAS]{\includegraphics[width=0.48\linewidth]{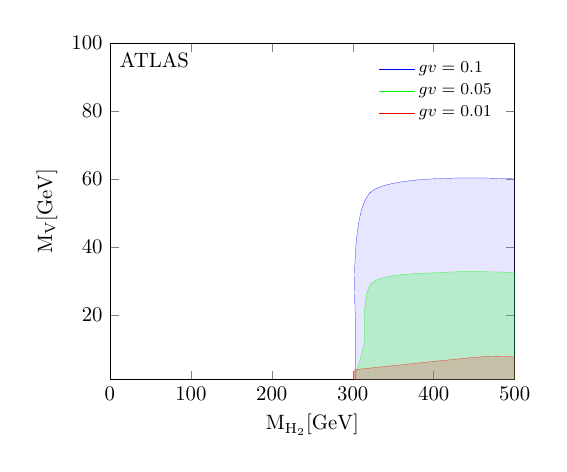}}
	\caption{The colored regions are excluded at 95\% CL by CMS (a) and ATLAS (b) measurements in $pp \to H_1VV,(H_1\to b\bar{b})$ channel for different $g_v$ values.}
	\label{Hbb_CMS}
\end{figure}

Figure \ref{Hbb_CMS}(a) shows the excluded regions in 95\% CL in the ($M_V$, $M_{H_2}$) plane for different $g_v$ values using the CMS results.
A similar result by using the ATLAS measurements is shown in Fig.~\ref{Hbb_CMS}(b).

\begin{figure}[b!]
	\centering
\subfloat[CMS]{\includegraphics[width=0.48\linewidth]{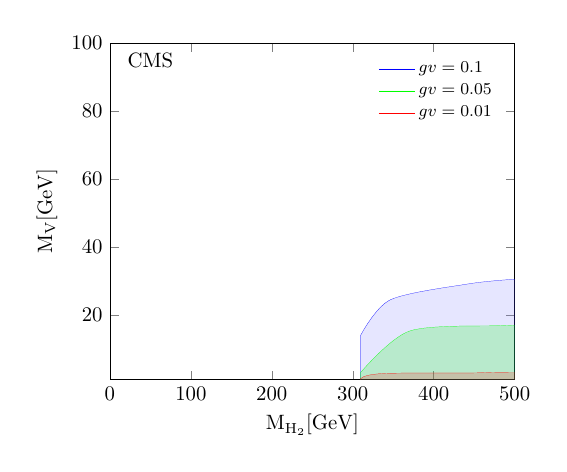}}
 \subfloat[ATLAS]{\includegraphics[width=0.48\linewidth]{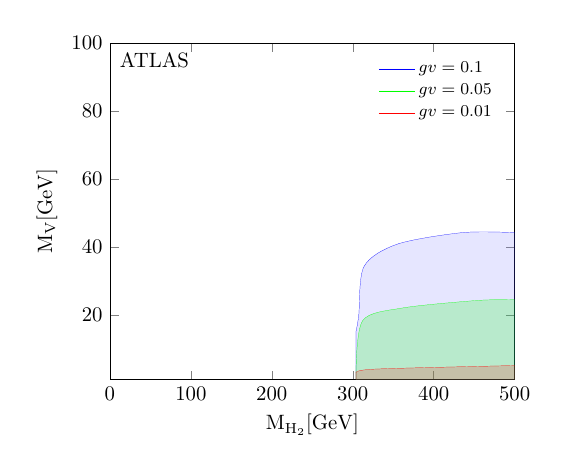}}
	\caption{The colored regions are excluded at 95\% CL by CMS (a) and ATLAS (b) measurements in $pp \to H_1VV,(H_1\to\tau^+\tau^-)$ channel for different $g_v$ values.}
	\label{H_CMS}
\end{figure}

In Ref.~\cite{CMS:2018nlv}, the CMS experiment considers the events consistent with the decay of the Higgs boson to two $\tau$ leptons to search for DM produced next to the Higgs boson. The analysis uses $35.9~\rm fb^{-1}$ of data in  $\sqrt{s}=13~\rm TeV$. Because of  different decay modes of the $\tau$ lepton, three final states are defined. Events with both $\tau$ leptons decay hadronically, or one of them decays hadronically and the other one decays to an electron or a muon, characterize these final states. No further binning is applied. The ATLAS experiment has reported the results of a similar search based on $139~\rm fb^{-1}$ of data in the same center-of-mass energy in Ref.~\cite{ATLAS:2023ild}. Only the hadronically decaying $\tau$ leptons are considered in this paper. The analysis defines two mass ranges. The high mass range is optimized for the booted Higgs bosons, but it is not completely separated from the low mass range. To avoid the complexities due to overlapping signal regions, only the low mass range is used in our analysis. Figure \ref{H_CMS}(a)
shows the excluded regions in 95\% CL in the ($M_V$, $M_{H_2}$) plane for different $g_v$ values using the CMS results.
A similar result by using the ATLAS measurements is shown in Fig.~\ref{H_CMS}(b). The reaches in $M_{H_2}$ direction are comparable, but ATLAS is more powerful in the $M_V$ direction. The ATLAS experiment uses more data, and this can be the source of this superiority.

\subsection{The LHC upper bounds for $pp \rightarrow t\bar{t}$+DM}
\begin{figure}
	\centering
	\begin{tikzpicture}
		\begin{feynman}
			\vertex (a);
			\vertex[below left of=a] (c) {$g$};
			\vertex [above right of=a] (f) ;
			\vertex [above left of=f] (b);
			\vertex[below right of=a] (e) {$\bar{t}$};
			\vertex[above left of =b] (d) {$g$};
			\vertex[above right of=b] (h) {$t$};
			\vertex[right of=f] (g);
			\vertex[below right of=g] (i) {$\bar{V}$};
			\vertex[above right of=g] (j) {$V$};
			\diagram{
				(a) -- [fermion](f);
				(f) --[fermion] (b);
				(d) --[boson] (b);
				(c) --[boson] (a);
				(e)--[fermion] (a);
				(f)--[scalar, edge label=$h_1/h_2$] (g);
				(i)--[fermion] (g)--[fermion] (j);
				(b)--[fermion] (h);
			};
		\end{feynman}
	\end{tikzpicture}
	\caption{Feynman diagrams relevant for the production of $t\bar{t}$ in association with DM at the LHC in the process pp $\to t\bar{t}$VV.}
	\label{diag.ttVV}
\end{figure}
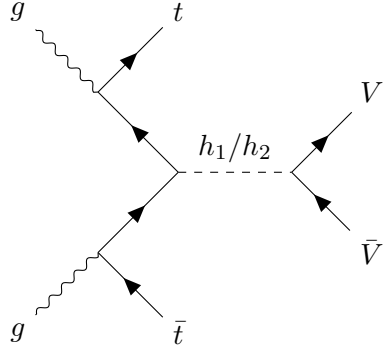

\begin{figure}[h!]
\centering
\subfloat[CMS]{\includegraphics[width=0.48\linewidth]{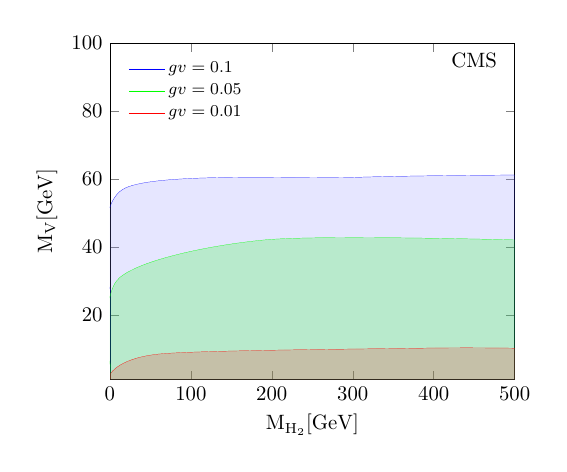}}
 \subfloat[ATLAS]{\includegraphics[width=0.48\linewidth]{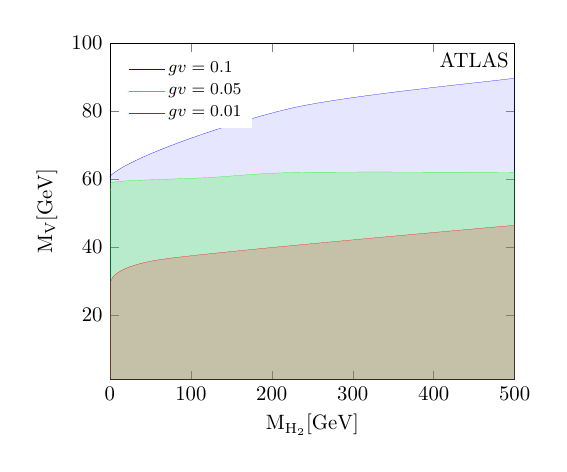}}
\caption{The colored regions are excluded at 95\% CL by CMS (a) and ATLAS (b)  measurements in $pp \to t\bar{t}VV$ channel for different $g_v$ values.}
\label{ttbar_CMS}
\end{figure}

Associate production of DM and top quark pair ($t\bar{t}$) can happen in Feynman diagrams like Fig.~\ref{diag.ttVV}.
The LHC experiments have reported several analyses on the search for DM production in association with $t\bar{t}$. In this paper, only the latest analyses are considered. The CMS experiment has used 2.2 fb$^{-1}$ of data in $\sqrt{s}=13~\rm TeV$ for this search~\cite{CMS:2017dcx}. The analysis combines the results from different $t\bar{t}$ final states, including zero, one or both top quarks decay hadronically. In all hadronic $t\bar{t}$ events, a special discriminator is used to categorize the events based on the number of the resolved top tagged. It helps to include in the analysis, the boosted top quarks which may have some overlaps between the jets. On the other hand, the ATLAS experiment has used 36.1 fb$^{-1}$ of data in the same center-of-mass energy~\cite{ATLAS:2017hoo}. They have presented the results for three sets of selection criteria when only fully hadronic and fully leptonic decays of $t\bar{t}$ are considered. They also have special treatment for highly boosted top quarks, by using the large-radius jets.
Figure \ref{ttbar_CMS}(a) shows the parts of ($M_V$, $M_{H_2}$) parameter space, excluded at 95\%  CL using CMS results when different $g_v$ values are assumed. The same result by using the ATLAS measurements is shown in Fig.~\ref{ttbar_CMS}(b).
Similar to the previous section, ATLAS uses more data and is more powerful in the $M_V$ direction.  Comparing to Figs.~\ref{Hbb_CMS} and \ref{H_CMS}, this channel is more powerful in  the $M_{H_2}$ direction. Our analysis using this channel can exclude all $M_{H_2}$ values for low $M_V$, which is a significant achievement.

\subsection{LHC constraints on $Z$+DM}
\begin{figure}[h!]
\centering
\subfloat[CMS]{\includegraphics[width=0.48\linewidth]{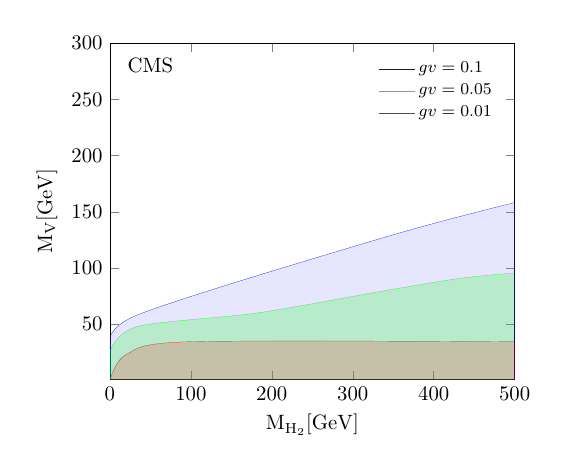}}
 \subfloat[ATLAS]{\includegraphics[width=0.48\linewidth]{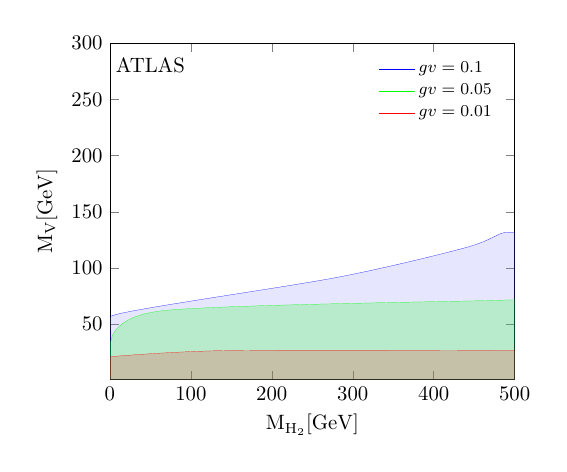}}
\caption{The colored regions are excluded at 95\% CL by CMS (a) and ATLAS (b) measurements in $pp \to ZVV,(Z\to \ell^+\ell^-)$  channel for different $g_v$ values.}
\label{z_CMS}
\end{figure}

The Feynman diagrams like Fig.~\ref{diag.XVV} can contribute in associate production of the $Z$ boson and DM.
The CMS experiment has reported the event yields of the backgrounds and data after applying the cuts in events with a $Z$ boson and the missing transverse momentum \cite{CMS:2020ulv}. The analysis uses 137 fb$^{-1}$ of data in $\sqrt{s}$=13 TeV for the search for DM produced in association with a leptonically decaying $Z$ boson. The results for dielectron and dimuon final states are summed up and reported in two jet multiplicity, zero and one jet, categories. Moreover, the ATLAS experiment has reported results on the same channel by applying different selection cuts \cite{ATLAS:2017nyv}. The analysis uses 36.1 fb$^{-1}$ of data in the center-of-mass energy of 13 TeV. The results are reported for dimuon and dielectron final states separately, and no further binning is done. Figure \ref{z_CMS}(a)
shows the parts of ($M_V$, $M_{H_2}$) parameter space, excluded at 95\%  CL using CMS results when different $g_v$ values are assumed. The same result by using the ATLAS measurements is shown in Fig.~\ref{z_CMS}(b). Two experiments have similar reaches in low $M_{H_2}$, but the CMS experiment, which uses more data in this channel, has a better exclusion for $M_V$ in high $M_{H_2}$. Comparing with other LHC constraints, this channel gives the best exclusion. It can increase the exclusion in both $M_V$ and $M_{H_2}$ directions comparing to DD results shown in Fig.~\ref{direct detection}.

\section{LHC constraints on  vector dark matter model with a $95~\rm GeV$ light Higgs} \label{sec.lhc95}
\begin{figure}[h!]
	\centering
	\subfloat[CMS]{\includegraphics[width=0.48\linewidth]{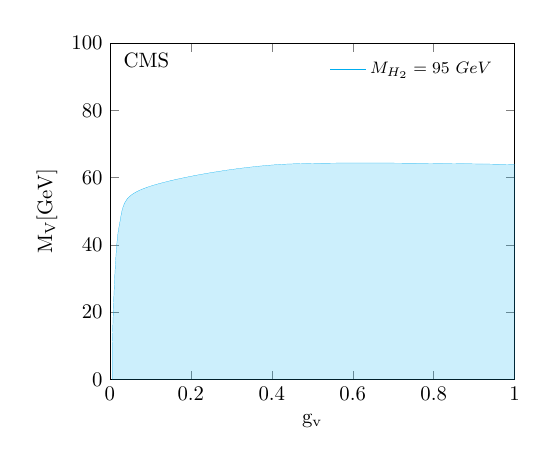}}
	\subfloat[ATLAS]{\includegraphics[width=0.48\linewidth]{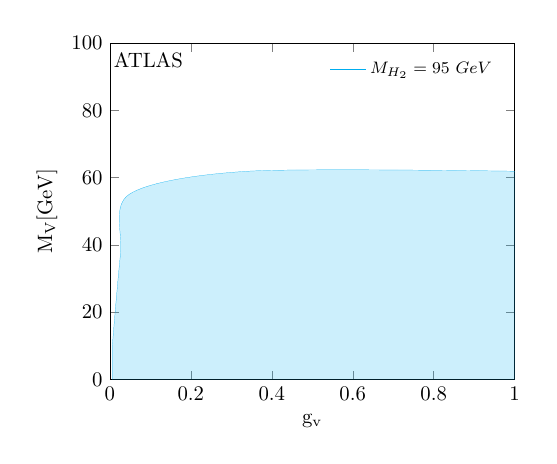}}
	\caption{The colored regions are excluded at 95\% CL by CMS (a) and ATLAS (b) measurements in $pp \to ZVV,(Z\to\ell^+\ell^-)$ channel for different $g_v$ values. The mass of $H_2$ is fixed to 95 GeV.}
	\label{H95Zll}
\end{figure}

\begin{figure}[h!]
	\centering
	\subfloat[CMS]{\includegraphics[width=0.48\linewidth]{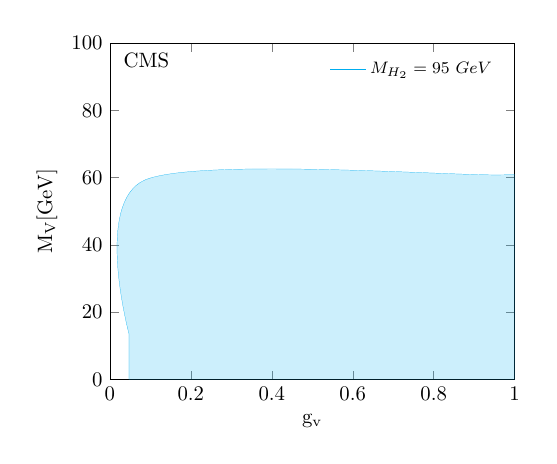}}
	\subfloat[ATLAS]{\includegraphics[width=0.48\linewidth]{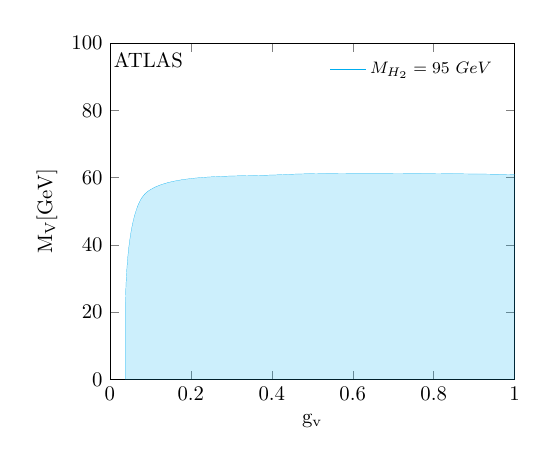}}
	\caption{The colored regions are excluded at 95\% CL by CMS (a) and ATLAS (b)  measurements in $pp \to t\bar{t}VV$ channel for different $g_v$ values. The mass of $H_2$ is fixed to 95 GeV.}
	\label{H95tt}
\end{figure}

Recently, the CMS Collaboration has reported new results, indicating a signal excess with a local (global) significance of $2.9\sigma$ ($1.3\sigma$) in the diphoton final state around $95~\rm GeV$, using all LHC data collected at $\sqrt{s} = 13~\rm TeV$ \cite{CMS:2024yhz}. The ATLAS
group also has reported a similar excess with a local significance of $1.7\sigma$ \cite{ATLAS:2024bjr}. Further anomalies in the $\tau\tau$ \cite{CMS:2022goy} and $b\bar{b}$ \cite{LEPWorkingGroupforHiggsbosonsearches:2003ing} final states have been reported by the CMS and LEP Collaborations, respectively. The impact of these reports has been studied in a variety of different BSMs\cite{Belyaev:2023xnv, Chen:2023bqr, Kalinowski:2024oyy, Ellwanger:2024txc, Kalinowski:2024uxe, Liu:2024cbr, Cao:2019ofo, Cao:2023gkc, Cao:2024axg, Ashanujjaman:2023etj, Borah:2023hqw, Banik:2023ecr, Bhattacharya:2023lmu, Benbrik:2024ptw, Diaz:2024yfu, Arhrib:2024wjj, Ge:2024rdr, Ahriche:2023hho, Aguilar-Saavedra:2023tql, Ellwanger:2024vvs, Wang:2024bkg, Arcadi:2023smv, Azevedo:2023zkg, Escribano:2023hxj, Biekotter:2023oen, Dutta:2023cig, Li:2022etb, Butterworth:2023rnw, Dev:2023kzu}. In the proposed model, the new Higgs emerges as a promising candidate. In this regard, $H_2$ is assumed to be the newly observed particle, and  $m_{H_2}=95~\rm GeV$. Looking at Figs.~\ref{Hbb_CMS}-\ref{z_CMS}, shows only  the processes $pp \to t\bar{t}VV$ and  $pp \to ZVV,(Z\to \ell^+\ell^-)$ have sensitivity in $m_{H_2}$ close to $95~\rm GeV$. Therefore the results of these channels are used to constrain the proposed model in the light of the newly observed 95 GeV excess. Figures \ref{H95Zll} and \ref{H95tt} show the excluded regions in $M_V$-$g_v$ plane, for  $pp \to ZVV,(Z\to \ell^+\ell^-)$ and $pp \to t\bar{t}VV$, respectively. The results based on the analysis of both CMS and ATLAS Collaborations are shown. Apart from very low values of $g_v$ for both channels and both collaborations, $M_V$ values below ~60 GeV are excluded. Dependence on $M_V$ is small but, in low values of $g_v$, dependence on this parameter is large, and small changes in coupling can switch between allowed and excluded regions. Similar behavior was seen in the previous shapes (Figs. \ref{Hbb_CMS}-\ref{z_CMS}) also, when the sensitive area is for $g_v$ between 0.01 and 0.1 and beyond this range, no sensitivity is seen.


It is necessary to mention that we consider the model in the leading order and it is impossible to study the decay of $H_2$ to diphoton. Looking at Fig.~\ref{feynman loop}, it can be seen that $H_2$ can decay to diphoton when loop corrections are considered. In our study, we only check, if  $H_2$  is responsible for the reported diphoton excess, which part of the parameter space is preferred. Checking the complete consistency between  $H_2$ of the model with the reported excess needs more details from the experiments (data and background yields in the vicinity of 95 GeV), which is not provided in the CMS and ATLAS papers or a complete analysis with full detector simulation and reconstruction of signal and backgrounds and comparison with LHC data. The former part is not accessible to the public yet; furthermore, the latter part is beyond the scope of this analysis.

\section{The H.E.S.S. upper bound } \label{sec.hess}
\begin{figure}
	\begin{center}
		\centerline{\hspace{0cm}\epsfig{figure=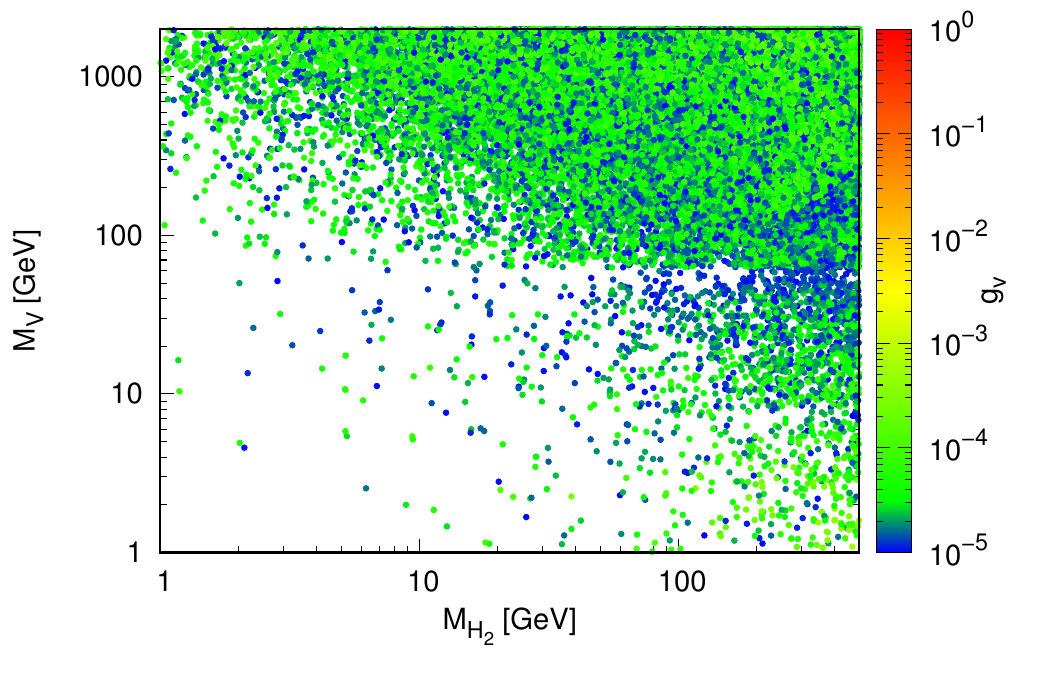,width=10cm}}
		\caption{The allowed range of parameter space consistent with H.E.S.S. data.} \label{HESS}
	\end{center}
\end{figure}

Indirect detection experiments of DM can be an attractive avenue for signs of DM. Photons, neutrinos, and positrons are the most important products from the annihilation of DM, among which photons are the best option due to their larger cross section and clearer signals. The H.E.S.S. experiment checks the cosmic $\gamma$-rays in the photon energy range of 0.03--100 TeV so it can observe high energy processes in the Universe\cite{HESS:2018cbt}. The monochromatic lines of photons are one of the promising messengers in the indirect search for DM. In our model, DM can annihilate to two photons through quantum loop corrections. The dominant Feynman diagrams contributing to this process are shown in Fig.~\ref{feynman loop}. The cross section is as follows:

\begin{figure}
	\begin{center}
		\centerline{\hspace{0cm}\epsfig{figure=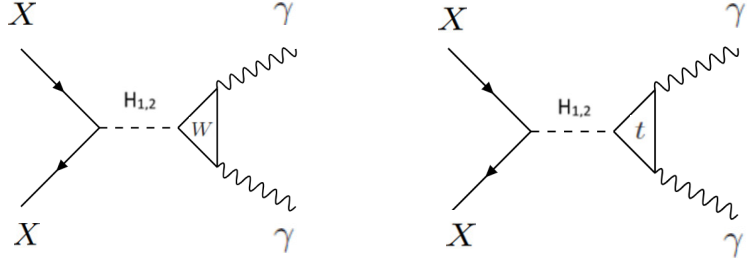,width=10cm}}
		\centerline{\vspace{-0.2cm}}
		\caption{The dominant Feynman diagrams for the annihilation of a DM pair into monochromatic $\gamma$-ray lines.} \label{feynman loop}
	\end{center}
\end{figure}

\begin{equation}
\sigma(s)\nu = \frac{1}{4\pi s}|M_{XX \longrightarrow\gamma\gamma}|^2,
\end{equation}
where $\sqrt{s}$ is the center-of-mass energy. The amplitude $M_{XX \longrightarrow\gamma\gamma}$ is given by
\begin{equation}
M_{XX \longrightarrow\gamma\gamma}(s)= i g_{v}^2 \nu _2 [\frac{i \sin \theta}{s-m_{H_1}^2 -im_{H_1}\Gamma_{H_1}}M_{{H_1}\longrightarrow\gamma\gamma}(s)+\frac{i \cos \theta}{s-m_{H_2}^2 -im_{H_2}\Gamma_{H_2}}M_{{H_2}\longrightarrow\gamma\gamma}(s)],
\end{equation}
where
\begin{equation}
M_{{H_1}\longrightarrow\gamma\gamma}=\frac{\alpha g_e s \cos \theta}{8\pi M_W}[3(\frac{2}{3})^2 F_t +F_W],
\end{equation}
\begin{equation}
M_{{H_2}\longrightarrow\gamma\gamma}=\frac{\alpha g_e s \sin \theta}{8\pi M_W}[3(\frac{2}{3})^2 F_t +F_W].
\end{equation}
Here, $g_e=2\sqrt{\pi\alpha}/sin \theta_w$ and
\begin{equation}
F_t = -2\tau [1+(1-\tau)f(\tau)],
\end{equation}
\begin{equation}
F_W= 2+3\tau+3\tau(2-\tau)f(\tau),
\end{equation}
where $\tau=\frac{4m_{i}^2}{s}$ with $i=t,W$ and
\begin{eqnarray}
f(\tau)=\bigg\lbrace \begin{array}{cc}
{(\sin^{-1} \sqrt{\frac{1}{\tau}})^2}&{if ~~~ \tau\geq1}\\{\frac{-1}{4}(ln \frac{1+\sqrt{1-\tau}}{1-\sqrt{1+\tau}}-i\pi)^2}&{if  ~~~ \tau<1}
\end{array}.
\end{eqnarray}
Therefore, the annihilation cross section of a pair of DM into two photons is as follows:
\begin{equation}
\sigma(s)\nu = \frac{g_{v}^4 \nu _2^2 }{4 \pi s } |\frac{\sin \theta M_{{H_1}\longrightarrow\gamma\gamma}}{s-m_{H_1}^2 -im_{H_1}\Gamma_{H_1}}+\frac{\cos \theta M_{{H_2}\longrightarrow\gamma\gamma}}{s-m_{H_2}^2 -im_{H_2}\Gamma_{H_2}}|^2.
\end{equation}
Finally, the average thermal cross section is equal to
\begin{equation}
\langle\sigma(s)\nu \rangle= \frac{1}{8M_V^4 T K_2^2 (\frac{M_X}{T})}\int_{4M_V^2}^{\infty}ds (s-4M_V^2) \sqrt{s} K_1 (\frac{\sqrt{s}}{T})\sigma(s),
\end{equation}
where $K_1$ and $K_2$ are modified Bessel functions and $T$ is the freeze-out temperature. Figure \ref{HESS} shows the allowed parameter space in agreement with the H.E.S.S. experiment. From the comparison of Figs \ref{HESS} and \ref{Relic}, it is clear that only for $1 \leq M_V \leq 10$ and $2 \leq M_{H_2} \leq 40 ~\rm GeV$ is the parameter space of our model is compatible with the H.E.S.S. results. Comparing this figure with relic density and direct detection results indicates that allowed  range of parameters space, which can simultaneously explain relic density, direct detection, and the H.E.S.S. experiment, is fairly small for the $g_v$ coupling constant.

\section{Conclusions} \label{sec.con}
In this paper, we revisit an extension of the SM by an extra $U_D(1)$ symmetry. The model consists of an extra scalar as the Higgs portal and a vector gauge field as a DM candidate.
It is shown that the model can satisfy the observed cosmological DM abundance via the standard freeze-out mechanism and, at the same time, fulfill the theoretical constraints such as perturbativity, unitarity, and positivity of the potential and also experimental constraints such as invisible Higgs decay and DD bounds. In the following, we focus on associate production of Higgs boson, $Z$ gauge boson, and top quark with the VDM. The results of the LHC experiments on similar final states are used to constrain the parameter space of the model. A statistical method is used to find the maximum room for the events from this DM model in LHC data. It is shown that the results of the associate production of $Z$ gauge boson and DM can exclude some parts of parameters space that are still allowed by relic density and the DD searches. In light of the LHC results, this analysis excludes all $M_{H_2}$ values for the light VDM, which is an unprecedented result. Inspired by the newly observed light Higgs boson close to 95 GeV, the model is tested to determine if its new scalar is the light Higgs boson. The excluded parameter space  in light of the LHC constraints and this assumption is also reported. In the end, the constraints from the indirect searches are also examined for completeness.

\bibliography{References}

\providecommand{\href}[2]{#2}\begingroup\raggedright\begin{thebibliography}{10}

\bibitem{Bertone:2016nfn}
G.~Bertone and D.~Hooper, \emph{{History of dark matter}},
  \href{https://doi.org/10.1103/RevModPhys.90.045002}{\emph{Rev. Mod. Phys.}
  {\bfseries 90} (2018) 045002}
  [\href{https://arxiv.org/abs/1605.04909}{{\ttfamily 1605.04909}}].

\bibitem{Feng:2022rxt}
J.~L. Feng, \emph{{The WIMP paradigm: Theme and variations}},
  \href{https://doi.org/10.21468/SciPostPhysLectNotes.71}{\emph{SciPost Phys.
  Lect. Notes} {\bfseries 71} (2023) 1}
  [\href{https://arxiv.org/abs/2212.02479}{{\ttfamily 2212.02479}}].

\bibitem{Lebedev:2021xey}
O.~Lebedev, \emph{{The Higgs portal to cosmology}},
  \href{https://doi.org/10.1016/j.ppnp.2021.103881}{\emph{Prog. Part. Nucl.
  Phys.} {\bfseries 120} (2021) 103881}
  [\href{https://arxiv.org/abs/2104.03342}{{\ttfamily 2104.03342}}].

\bibitem{CMS:2024yhz}
{\scshape CMS} collaboration, \emph{{Search for a standard model-like Higgs
  boson in the mass range between 70 and 110 GeV in the diphoton final state in
  proton-proton collisions at $\sqrt{s}$ = 13 TeV}},
  \href{https://arxiv.org/abs/2405.18149}{{\ttfamily 2405.18149}}.

\bibitem{ATLAS:2024bjr}
{\scshape ATLAS} collaboration, \emph{{Search for diphoton resonances in the 66
  to 110 GeV mass range using $pp$ collisions at $\sqrt{s}=13$ TeV with the
  ATLAS detector}},  \href{https://arxiv.org/abs/2407.07546}{{\ttfamily
  2407.07546}}.

\bibitem{YaserAyazi:2022tbn}
S.~Yaser~Ayazi and M.~Hosseini, \emph{{W-boson mass anomaly and vacuum
  structure in vector dark matter model with a singlet scalar mediator}},
  \href{https://doi.org/10.1142/S0217751X2340002X}{\emph{Int. J. Mod. Phys. A}
  {\bfseries 38} (2023) 2340002}
  [\href{https://arxiv.org/abs/2206.11041}{{\ttfamily 2206.11041}}].

\bibitem{Glaus:2019itb}
S.~Glaus, M.~M\"uhlleitner, J.~M\"uller, S.~Patel and R.~Santos,
  \emph{{Electroweak Corrections to Dark Matter Direct Detection in a Vector
  Dark Matter Model}},
  \href{https://doi.org/10.1007/JHEP10(2019)152}{\emph{JHEP} {\bfseries 10}
  (2019) 152} [\href{https://arxiv.org/abs/1908.09249}{{\ttfamily
  1908.09249}}].

\bibitem{Hashino:2018zsi}
K.~Hashino, M.~Kakizaki, S.~Kanemura, P.~Ko and T.~Matsui, \emph{{Gravitational
  waves from first order electroweak phase transition in models with the
  U(1)$_{X}$ gauge symmetry}},
  \href{https://doi.org/10.1007/JHEP06(2018)088}{\emph{JHEP} {\bfseries 06}
  (2018) 088} [\href{https://arxiv.org/abs/1802.02947}{{\ttfamily
  1802.02947}}].

\bibitem{Amiri:2022cbv}
A.~Amiri, B.~D\'\i{}az~S\'aez and K.~Ghorbani, \emph{{(sub)GeV dark matter in
  the U(1)X Higgs portal model}},
  \href{https://doi.org/10.1016/j.physletb.2023.138119}{\emph{Phys. Lett. B}
  {\bfseries 844} (2023) 138119}
  [\href{https://arxiv.org/abs/2209.11723}{{\ttfamily 2209.11723}}].

\bibitem{Barducci:2016pcb}
D.~Barducci, G.~Belanger, J.~Bernon, F.~Boudjema, J.~Da~Silva, S.~Kraml et~al.,
  \emph{{Collider limits on new physics within micrOMEGAs$\_$4.3}},
  \href{https://doi.org/10.1016/j.cpc.2017.08.028}{\emph{Comput. Phys. Commun.}
  {\bfseries 222} (2018) 327}
  [\href{https://arxiv.org/abs/1606.03834}{{\ttfamily 1606.03834}}].

\bibitem{Planck:2018vyg}
{\scshape Planck} collaboration, \emph{{Planck 2018 results. VI. Cosmological
  parameters}},
  \href{https://doi.org/10.1051/0004-6361/201833910}{\emph{Astron. Astrophys.}
  {\bfseries 641} (2020) A6}
  [\href{https://arxiv.org/abs/1807.06209}{{\ttfamily 1807.06209}}].

\bibitem{DAgnolo:2015ujb}
R.~T. D'Agnolo and J.~T. Ruderman, \emph{{Light Dark Matter from Forbidden
  Channels}}, \href{https://doi.org/10.1103/PhysRevLett.115.061301}{\emph{Phys.
  Rev. Lett.} {\bfseries 115} (2015) 061301}
  [\href{https://arxiv.org/abs/1505.07107}{{\ttfamily 1505.07107}}].

\bibitem{Farzinnia:2013pga}
A.~Farzinnia, H.-J. He and J.~Ren, \emph{{Natural Electroweak Symmetry Breaking
  from Scale Invariant Higgs Mechanism}},
  \href{https://doi.org/10.1016/j.physletb.2013.09.060}{\emph{Phys. Lett. B}
  {\bfseries 727} (2013) 141}
  [\href{https://arxiv.org/abs/1308.0295}{{\ttfamily 1308.0295}}].

\bibitem{Farzinnia:2014xia}
A.~Farzinnia and J.~Ren, \emph{{Higgs Partner Searches and Dark Matter
  Phenomenology in a Classically Scale Invariant Higgs Boson Sector}},
  \href{https://doi.org/10.1103/PhysRevD.90.015019}{\emph{Phys. Rev. D}
  {\bfseries 90} (2014) 015019}
  [\href{https://arxiv.org/abs/1405.0498}{{\ttfamily 1405.0498}}].

\bibitem{YaserAyazi:2019caf}
S.~Yaser~Ayazi and A.~Mohamadnejad, \emph{{Conformal vector dark matter and
  strongly first-order electroweak phase transition}},
  \href{https://doi.org/10.1007/JHEP03(2019)181}{\emph{JHEP} {\bfseries 03}
  (2019) 181} [\href{https://arxiv.org/abs/1901.04168}{{\ttfamily
  1901.04168}}].

\bibitem{XENON:2023cxc}
{\scshape XENON} collaboration, \emph{{First Dark Matter Search with Nuclear
  Recoils from the XENONnT Experiment}},
  \href{https://doi.org/10.1103/PhysRevLett.131.041003}{\emph{Phys. Rev. Lett.}
  {\bfseries 131} (2023) 041003}
  [\href{https://arxiv.org/abs/2303.14729}{{\ttfamily 2303.14729}}].

\bibitem{LZ:2022lsv}
{\scshape LZ} collaboration, \emph{{First Dark Matter Search Results from the
  LUX-ZEPLIN (LZ) Experiment}},
  \href{https://doi.org/10.1103/PhysRevLett.131.041002}{\emph{Phys. Rev. Lett.}
  {\bfseries 131} (2023) 041002}
  [\href{https://arxiv.org/abs/2207.03764}{{\ttfamily 2207.03764}}].

\bibitem{XENON:2019gfn}
{\scshape XENON} collaboration, \emph{{Light Dark Matter Search with Ionization
  Signals in XENON1T}},
  \href{https://doi.org/10.1103/PhysRevLett.123.251801}{\emph{Phys. Rev. Lett.}
  {\bfseries 123} (2019) 251801}
  [\href{https://arxiv.org/abs/1907.11485}{{\ttfamily 1907.11485}}].

\bibitem{Brun:1997pa}
R.~Brun and F.~Rademakers, \emph{{ROOT: An object oriented data analysis
  framework}}, \href{https://doi.org/10.1016/S0168-9002(97)00048-X}{\emph{Nucl.
  Instrum. Meth. A} {\bfseries 389} (1997) 81}.

\bibitem{Cousins:1991qz}
R.~D. Cousins and V.~L. Highland, \emph{{Incorporating systematic uncertainties
  into an upper limit}},
  \href{https://doi.org/10.1016/0168-9002(92)90794-5}{\emph{Nucl. Instrum.
  Meth. A} {\bfseries 320} (1992) 331}.

\bibitem{Conrad:2002kn}
J.~Conrad, O.~Botner, A.~Hallgren and C.~Perez de~los Heros, \emph{{Including
  systematic uncertainties in confidence interval construction for Poisson
  statistics}}, \href{https://doi.org/10.1103/PhysRevD.67.012002}{\emph{Phys.
  Rev. D} {\bfseries 67} (2003) 012002}
  [\href{https://arxiv.org/abs/hep-ex/0202013}{{\ttfamily hep-ex/0202013}}].

\bibitem{PaktinatMehdiabadi:2017clu}
S.~Paktinat~Mehdiabadi and L.~Zamiri, \emph{{W' pair production in the light of
  CMS searches}}, \href{https://doi.org/10.1088/1361-6471/aab415}{\emph{J.
  Phys. G} {\bfseries 45} (2018) 055004}
  [\href{https://arxiv.org/abs/1710.05153}{{\ttfamily 1710.05153}}].

\bibitem{Alwall_2014}
J.~Alwall, R.~Frederix, S.~Frixione, V.~Hirschi, F.~Maltoni, O.~Mattelaer
  et~al., \emph{The automated computation of tree-level and next-to-leading
  order differential cross sections, and their matching to parton shower
  simulations}, \href{https://doi.org/10.1007/jhep07(2014)079}{\emph{Journal of
  High Energy Physics} {\bfseries 2014} (2014) }.

\bibitem{Semenov:2014rea}
A.~Semenov, \emph{{LanHEP \textemdash{} A package for automatic generation of
  Feynman rules from the Lagrangian. Version 3.2}},
  \href{https://doi.org/10.1016/j.cpc.2016.01.003}{\emph{Comput. Phys. Commun.}
  {\bfseries 201} (2016) 167}
  [\href{https://arxiv.org/abs/1412.5016}{{\ttfamily 1412.5016}}].

\bibitem{Bierlich:2022pfr}
C.~Bierlich et~al., \emph{{A comprehensive guide to the physics and usage of
  PYTHIA 8.3}},
  \href{https://doi.org/10.21468/SciPostPhysCodeb.8}{\emph{SciPost Phys.
  Codeb.} {\bfseries 2022} (2022) 8}
  [\href{https://arxiv.org/abs/2203.11601}{{\ttfamily 2203.11601}}].

\bibitem{CMS:2018zjv}
{\scshape CMS} collaboration, \emph{{Search for dark matter produced in
  association with a Higgs boson decaying to a pair of bottom quarks in
  proton\textendash{}proton collisions at $\sqrt{s}=13\,\text {Te}\text {V}
  $}}, \href{https://doi.org/10.1140/epjc/s10052-019-6730-7}{\emph{Eur. Phys.
  J. C} {\bfseries 79} (2019) 280}
  [\href{https://arxiv.org/abs/1811.06562}{{\ttfamily 1811.06562}}].

\bibitem{ATLAS:2021shl}
{\scshape ATLAS} collaboration, \emph{{Search for dark matter produced in
  association with a Standard Model Higgs boson decaying into b-quarks using
  the full Run 2 dataset from the ATLAS detector}},
  \href{https://doi.org/10.1007/JHEP11(2021)209}{\emph{JHEP} {\bfseries 11}
  (2021) 209} [\href{https://arxiv.org/abs/2108.13391}{{\ttfamily
  2108.13391}}].

\bibitem{CMS:2018nlv}
{\scshape CMS} collaboration, \emph{{Search for dark matter produced in
  association with a Higgs boson decaying to $\gamma\gamma$ or $\tau^+\tau^-$
  at $\sqrt{s} =$ 13 TeV}},
  \href{https://doi.org/10.1007/JHEP09(2018)046}{\emph{JHEP} {\bfseries 09}
  (2018) 046} [\href{https://arxiv.org/abs/1806.04771}{{\ttfamily
  1806.04771}}].

\bibitem{ATLAS:2023ild}
{\scshape ATLAS} collaboration, \emph{{Search for dark matter produced in
  association with a Higgs boson decaying to tau leptons at $ \sqrt{s} $ = 13
  TeV with the ATLAS detector}},
  \href{https://doi.org/10.1007/JHEP09(2023)189}{\emph{JHEP} {\bfseries 09}
  (2023) 189} [\href{https://arxiv.org/abs/2305.12938}{{\ttfamily
  2305.12938}}].

\bibitem{CMS:2017dcx}
{\scshape CMS} collaboration, \emph{{Search for dark matter produced in
  association with heavy-flavor quark pairs in proton-proton collisions at
  $\sqrt{s}=13$ TeV}},
  \href{https://doi.org/10.1140/epjc/s10052-017-5317-4}{\emph{Eur. Phys. J. C}
  {\bfseries 77} (2017) 845}
  [\href{https://arxiv.org/abs/1706.02581}{{\ttfamily 1706.02581}}].

\bibitem{ATLAS:2017hoo}
{\scshape ATLAS} collaboration, \emph{{Search for dark matter produced in
  association with bottom or top quarks in $\sqrt{s}=13$ TeV pp collisions with
  the ATLAS detector}},
  \href{https://doi.org/10.1140/epjc/s10052-017-5486-1}{\emph{Eur. Phys. J. C}
  {\bfseries 78} (2018) 18} [\href{https://arxiv.org/abs/1710.11412}{{\ttfamily
  1710.11412}}].

\bibitem{CMS:2020ulv}
{\scshape CMS} collaboration, \emph{{Search for dark matter produced in
  association with a leptonically decaying Z boson in proton-proton collisions
  at $\sqrt{s} =$ 13 TeV}},
  \href{https://doi.org/10.1140/epjc/s10052-020-08739-5}{\emph{Eur. Phys. J. C}
  {\bfseries 81} (2021) 13} [\href{https://arxiv.org/abs/2008.04735}{{\ttfamily
  2008.04735}}].

\bibitem{ATLAS:2017nyv}
{\scshape ATLAS} collaboration, \emph{{Search for an invisibly decaying Higgs
  boson or dark matter candidates produced in association with a $Z$ boson in
  $pp$ collisions at $\sqrt{s} =$ 13 TeV with the ATLAS detector}},
  \href{https://doi.org/10.1016/j.physletb.2017.11.049}{\emph{Phys. Lett. B}
  {\bfseries 776} (2018) 318}
  [\href{https://arxiv.org/abs/1708.09624}{{\ttfamily 1708.09624}}].

\bibitem{CMS:2022goy}
{\scshape CMS} collaboration, \emph{{Searches for additional Higgs bosons and
  for vector leptoquarks in $\tau\tau$ final states in proton-proton collisions
  at $\sqrt{s}$ = 13 TeV}},
  \href{https://doi.org/10.1007/JHEP07(2023)073}{\emph{JHEP} {\bfseries 07}
  (2023) 073} [\href{https://arxiv.org/abs/2208.02717}{{\ttfamily
  2208.02717}}].

\bibitem{LEPWorkingGroupforHiggsbosonsearches:2003ing}
{\scshape LEP Working Group for Higgs boson searches, ALEPH, DELPHI, L3, OPAL}
  collaboration, \emph{{Search for the standard model Higgs boson at LEP}},
  \href{https://doi.org/10.1016/S0370-2693(03)00614-2}{\emph{Phys. Lett. B}
  {\bfseries 565} (2003) 61}
  [\href{https://arxiv.org/abs/hep-ex/0306033}{{\ttfamily hep-ex/0306033}}].

\bibitem{Belyaev:2023xnv}
A.~Belyaev, R.~Benbrik, M.~Boukidi, M.~Chakraborti, S.~Moretti and S.~Semlali,
  \emph{{Explanation of the hints for a 95 GeV Higgs boson within a 2-Higgs
  Doublet Model}}, \href{https://doi.org/10.1007/JHEP05(2024)209}{\emph{JHEP}
  {\bfseries 05} (2024) 209}
  [\href{https://arxiv.org/abs/2306.09029}{{\ttfamily 2306.09029}}].

\bibitem{Chen:2023bqr}
T.-K. Chen, C.-W. Chiang, S.~Heinemeyer and G.~Weiglein, \emph{{95~GeV Higgs
  boson in the Georgi-Machacek model}},
  \href{https://doi.org/10.1103/PhysRevD.109.075043}{\emph{Phys. Rev. D}
  {\bfseries 109} (2024) 075043}
  [\href{https://arxiv.org/abs/2312.13239}{{\ttfamily 2312.13239}}].

\bibitem{Kalinowski:2024oyy}
J.~Kalinowski and W.~Kotlarski, \emph{{Constraining Higgs sectors of BSM models
  -- the case of 95 GeV ''Higgs''}},  in \emph{{Workshop on the Standard Model
  and Beyond}}, 4, 2024, \href{https://arxiv.org/abs/2404.12233}{{\ttfamily
  2404.12233}}.

\bibitem{Ellwanger:2024txc}
U.~Ellwanger and C.~Hugonie, \emph{{Nmssm with correct relic density and an
  additional 95~GeV Higgs boson}},
  \href{https://doi.org/10.1140/epjc/s10052-024-12886-4}{\emph{Eur. Phys. J. C}
  {\bfseries 84} (2024) 526}
  [\href{https://arxiv.org/abs/2403.16884}{{\ttfamily 2403.16884}}].

\bibitem{Kalinowski:2024uxe}
J.~Kalinowski and W.~Kotlarski, \emph{{Interpreting 95 GeV di-photon/$
  b\overline{b} $ excesses as a lightest Higgs boson of the MRSSM}},
  \href{https://doi.org/10.1007/JHEP07(2024)037}{\emph{JHEP} {\bfseries 07}
  (2024) 037} [\href{https://arxiv.org/abs/2403.08720}{{\ttfamily
  2403.08720}}].

\bibitem{Liu:2024cbr}
C.-X. Liu, Y.~Zhou, X.-Y. Zheng, J.~Ma, T.-F. Feng and H.-B. Zhang,
  \emph{{95~GeV excess in a CP-violating \ensuremath{\mu}-from-\ensuremath{\nu}
  SSM}}, \href{https://doi.org/10.1103/PhysRevD.109.056001}{\emph{Phys. Rev. D}
  {\bfseries 109} (2024) 056001}
  [\href{https://arxiv.org/abs/2402.00727}{{\ttfamily 2402.00727}}].

\bibitem{Cao:2019ofo}
J.~Cao, X.~Jia, Y.~Yue, H.~Zhou and P.~Zhu, \emph{{96 GeV diphoton excess in
  seesaw extensions of the natural NMSSM}},
  \href{https://doi.org/10.1103/PhysRevD.101.055008}{\emph{Phys. Rev. D}
  {\bfseries 101} (2020) 055008}
  [\href{https://arxiv.org/abs/1908.07206}{{\ttfamily 1908.07206}}].

\bibitem{Cao:2023gkc}
J.~Cao, X.~Jia, J.~Lian and L.~Meng, \emph{{95~GeV diphoton and
  bb\textasciimacron{} excesses in the general next-to-minimal supersymmetric
  standard model}},
  \href{https://doi.org/10.1103/PhysRevD.109.075001}{\emph{Phys. Rev. D}
  {\bfseries 109} (2024) 075001}
  [\href{https://arxiv.org/abs/2310.08436}{{\ttfamily 2310.08436}}].

\bibitem{Cao:2024axg}
J.~Cao, X.~Jia and J.~Lian, \emph{{Unified Interpretation of Muon g-2 anomaly,
  95 GeV Diphoton, and $b\bar{b}$ Excesses in the General Next-to-Minimal
  Supersymmetric Standard Model}},
  \href{https://arxiv.org/abs/2402.15847}{{\ttfamily 2402.15847}}.

\bibitem{Ashanujjaman:2023etj}
S.~Ashanujjaman, S.~Banik, G.~Coloretti, A.~Crivellin, B.~Mellado and A.-T.
  Mulaudzi, \emph{{SU(2)L triplet scalar as the origin of the 95~GeV excess?}},
  \href{https://doi.org/10.1103/PhysRevD.108.L091704}{\emph{Phys. Rev. D}
  {\bfseries 108} (2023) L091704}
  [\href{https://arxiv.org/abs/2306.15722}{{\ttfamily 2306.15722}}].

\bibitem{Borah:2023hqw}
D.~Borah, S.~Mahapatra, P.~K. Paul and N.~Sahu, \emph{{Scotogenic
  U(1)L\ensuremath{\mu}-L\ensuremath{\tau} origin of (g-2)\ensuremath{\mu},
  W-mass anomaly and 95~GeV excess}},
  \href{https://doi.org/10.1103/PhysRevD.109.055021}{\emph{Phys. Rev. D}
  {\bfseries 109} (2024) 055021}
  [\href{https://arxiv.org/abs/2310.11953}{{\ttfamily 2310.11953}}].

\bibitem{Banik:2023ecr}
S.~Banik, A.~Crivellin, S.~Iguro and T.~Kitahara, \emph{{Asymmetric di-Higgs
  signals of the next-to-minimal 2HDM with a U(1) symmetry}},
  \href{https://doi.org/10.1103/PhysRevD.108.075011}{\emph{Phys. Rev. D}
  {\bfseries 108} (2023) 075011}
  [\href{https://arxiv.org/abs/2303.11351}{{\ttfamily 2303.11351}}].

\bibitem{Bhattacharya:2023lmu}
S.~Bhattacharya, G.~Coloretti, A.~Crivellin, S.-E. Dahbi, Y.~Fang, M.~Kumar
  et~al., \emph{{Growing Excesses of New Scalars at the Electroweak Scale}},
  \href{https://arxiv.org/abs/2306.17209}{{\ttfamily 2306.17209}}.

\bibitem{Benbrik:2024ptw}
R.~Benbrik, M.~Boukidi and S.~Moretti, \emph{{Superposition of CP-Even and
  CP-Odd Higgs Resonances: Explaining the 95 GeV Excesses within a Two-Higgs
  Doublet Model}},  \href{https://arxiv.org/abs/2405.02899}{{\ttfamily
  2405.02899}}.

\bibitem{Diaz:2024yfu}
M.~A. Diaz, G.~Cerro, S.~Dasmahapatra and S.~Moretti, \emph{{Bayesian Active
  Search on Parameter Space: a 95 GeV Spin-0 Resonance in the ($B-L$)SSM}},
  \href{https://arxiv.org/abs/2404.18653}{{\ttfamily 2404.18653}}.

\bibitem{Arhrib:2024wjj}
A.~Arhrib, K.~H. Phan, V.~Q. Tran and T.-C. Yuan, \emph{{When Standard Model
  Higgs Meets Its Lighter 95 GeV Higgs}},
  \href{https://arxiv.org/abs/2405.03127}{{\ttfamily 2405.03127}}.

\bibitem{Ge:2024rdr}
Z.-f. Ge, F.-Y. Niu and J.-L. Yang, \emph{{The origin of the 95~GeV excess in
  the flavor-dependent $U(1)_X$ model}},
  \href{https://doi.org/10.1140/epjc/s10052-024-12872-w}{\emph{Eur. Phys. J. C}
  {\bfseries 84} (2024) 548}
  [\href{https://arxiv.org/abs/2405.07243}{{\ttfamily 2405.07243}}].

\bibitem{Ahriche:2023hho}
A.~Ahriche, M.~L. Bellilet, M.~O. Khojali, M.~Kumar and A.-T. Mulaudzi,
  \emph{{Scale invariant scotogenic model: CDF-II W-boson mass and the 95~GeV
  excesses}}, \href{https://doi.org/10.1103/PhysRevD.110.015025}{\emph{Phys.
  Rev. D} {\bfseries 110} (2024) 015025}
  [\href{https://arxiv.org/abs/2311.08297}{{\ttfamily 2311.08297}}].

\bibitem{Aguilar-Saavedra:2023tql}
J.~A. Aguilar-Saavedra, H.~B. C\^amara, F.~R. Joaquim and J.~F. Seabra,
  \emph{{Confronting the 95 GeV excesses within the U(1)'-extended
  next-to-minimal 2HDM}},
  \href{https://doi.org/10.1103/PhysRevD.108.075020}{\emph{Phys. Rev. D}
  {\bfseries 108} (2023) 075020}
  [\href{https://arxiv.org/abs/2307.03768}{{\ttfamily 2307.03768}}].

\bibitem{Ellwanger:2024vvs}
U.~Ellwanger, C.~Hugonie, S.~F. King and S.~Moretti, \emph{{NMSSM Explanation
  for Excesses in the Search for Neutralinos and Charginos and a 95 GeV Higgs
  Boson}},  \href{https://arxiv.org/abs/2404.19338}{{\ttfamily 2404.19338}}.

\bibitem{Wang:2024bkg}
K.~Wang and J.~Zhu, \emph{{95 GeV light Higgs in the top-pair-associated
  diphoton channel at the LHC in the minimal dilaton model*}},
  \href{https://doi.org/10.1088/1674-1137/ad4268}{\emph{Chin. Phys. C}
  {\bfseries 48} (2024) 073105}
  [\href{https://arxiv.org/abs/2402.11232}{{\ttfamily 2402.11232}}].

\bibitem{Arcadi:2023smv}
G.~Arcadi, G.~Busoni, D.~Cabo-Almeida and N.~Krishnan, \emph{{Is there a
  (Pseudo)Scalar at 95 GeV?}},
  \href{https://arxiv.org/abs/2311.14486}{{\ttfamily 2311.14486}}.

\bibitem{Azevedo:2023zkg}
D.~Azevedo, T.~Biek\"otter and P.~M. Ferreira, \emph{{2HDM interpretations of
  the CMS diphoton excess at 95 GeV}},
  \href{https://doi.org/10.1007/JHEP11(2023)017}{\emph{JHEP} {\bfseries 11}
  (2023) 017} [\href{https://arxiv.org/abs/2305.19716}{{\ttfamily
  2305.19716}}].

\bibitem{Escribano:2023hxj}
P.~Escribano, V.~M. Lozano and A.~Vicente, \emph{{Scotogenic explanation for
  the 95~GeV excesses}},
  \href{https://doi.org/10.1103/PhysRevD.108.115001}{\emph{Phys. Rev. D}
  {\bfseries 108} (2023) 115001}
  [\href{https://arxiv.org/abs/2306.03735}{{\ttfamily 2306.03735}}].

\bibitem{Biekotter:2023oen}
T.~Biek\"otter, S.~Heinemeyer and G.~Weiglein, \emph{{95.4~GeV diphoton excess
  at ATLAS and CMS}},
  \href{https://doi.org/10.1103/PhysRevD.109.035005}{\emph{Phys. Rev. D}
  {\bfseries 109} (2024) 035005}
  [\href{https://arxiv.org/abs/2306.03889}{{\ttfamily 2306.03889}}].

\bibitem{Dutta:2023cig}
J.~Dutta, J.~Lahiri, C.~Li, G.~Moortgat-Pick, S.~F. Tabira and J.~A. Ziegler,
  \emph{{Dark Matter Phenomenology in 2HDMS in light of the 95 GeV excess}},
  \href{https://arxiv.org/abs/2308.05653}{{\ttfamily 2308.05653}}.

\bibitem{Li:2022etb}
W.~Li, H.~Qiao and J.~Zhu, \emph{{Light Higgs boson in the NMSSM confronted
  with the CMS di-photon and di-tau excesses*}},
  \href{https://doi.org/10.1088/1674-1137/acfaf1}{\emph{Chin. Phys. C}
  {\bfseries 47} (2023) 123102}
  [\href{https://arxiv.org/abs/2212.11739}{{\ttfamily 2212.11739}}].

\bibitem{Butterworth:2023rnw}
J.~Butterworth, H.~Debnath, P.~Fileviez~Perez and F.~Mitchell, \emph{{Custodial
  symmetry breaking and Higgs boson signatures at the LHC}},
  \href{https://doi.org/10.1103/PhysRevD.109.095014}{\emph{Phys. Rev. D}
  {\bfseries 109} (2024) 095014}
  [\href{https://arxiv.org/abs/2309.10027}{{\ttfamily 2309.10027}}].

\bibitem{Dev:2023kzu}
P.~S.~B. Dev, R.~N. Mohapatra and Y.~Zhang, \emph{{Explanation of the 95 GeV
  \ensuremath{\gamma}\ensuremath{\gamma} and bb\textasciimacron{} excesses in
  the minimal left-right symmetric model}},
  \href{https://doi.org/10.1016/j.physletb.2024.138481}{\emph{Phys. Lett. B}
  {\bfseries 849} (2024) 138481}
  [\href{https://arxiv.org/abs/2312.17733}{{\ttfamily 2312.17733}}].

\bibitem{HESS:2018cbt}
{\scshape HESS} collaboration, \emph{{Search for $\gamma$-Ray Line Signals from
  Dark Matter Annihilations in the Inner Galactic Halo from 10 Years of
  Observations with H.E.S.S.}},
  \href{https://doi.org/10.1103/PhysRevLett.120.201101}{\emph{Phys. Rev. Lett.}
  {\bfseries 120} (2018) 201101}
  [\href{https://arxiv.org/abs/1805.05741}{{\ttfamily 1805.05741}}].

\end{thebibliography}\endgroup
\bibliographystyle{JHEP}

\end{document}